\documentstyle[prl,aps,multicol]{revtex}

\begin{document}

\title{ ADIABATIC MEASUREMENTS ON METASTABLE SYSTEMS}
\author{Y. Aharonov$^{a,b}$, S. Massar$^a$, S. Popescu$^c$, 
J. Tollaksen$^c$, and L. Vaidman$^a$
}

\address{$^a$ School of Physics and Astronomy,
Raymond and Beverly Sackler Faculty of Exact Sciences, \\
Tel-Aviv University, Tel-Aviv 69978, Israel.\\
$^b$ Physics Department, University of South Carolina,
Columbia, South Carolina 29208, USA\\
$^c$  Physics Department, Boston University, 
Boston, MA 02215, USA}

\date{}

\maketitle

\begin{abstract}
In several situations, most notably when describing metastable states,
a system can evolve according to an effective non hermitian Hamiltonian.
To each eigenvalue of a non hermitian Hamiltonian is
associated an eigenstate $\vert\phi\rangle$
which evolves forward in time and an eigenstate $\langle{\psi}\vert$ which
evolves backward in time. Quantum measurements on such systems are
analyzed
in detail with particular emphasis on adiabatic measurements in which 
the measuring device is coupled weakly to the system. It is 
shown that in this case 
the outcome of the measurement of an observable $A$ is   the weak  value
$\langle{\psi}\vert A\vert\phi\rangle / 
\langle{\psi}\vert{\phi}\rangle $  associated to the 
two-state  vector $\langle{\psi}\vert$ $\vert\phi\rangle$ 
corresponding to one of
the eigenvalues of the non hermitian Hamiltonian. The possibility of
performing such measurements in a laboratory is discussed.
\end{abstract}

\begin{multicols}{2}

Any interaction between two systems can be regarded, in a very wide
sense, as a ``measurement" since the state of one of the systems, the
measuring device, is affected by the state of the other one, the
measured system.  In general, however, this interaction is not very
``clean", that is, the information about the properties of the measured
system cannot be read easily from the final state of the measuring
device. Only some very particular classes of interactions are clean
enough and are called ``measurements" in the usual, more restricted,
sense.

The best known type of quantum measurement is the
von Neumann ideal measurement wherein the system is coupled 
impulsively to the measuring device. The Hamiltonian describing such a  
measurement is
\begin{equation}
H=H_0 + g(t) P A  + H_{MD} ,
\label{one}
\end{equation}
where $H_0$ is the free Hamiltonian of the system, $H_{MD}$ is the free 
Hamiltonian of the measuring
device,
$P$ is the momentum conjugate to the position variable 
$Q$ of the pointer 
of the measuring device, $A$ is the observable to be measured. The coupling
parameter $g(t)$ is normalized to $\int\!  g(t) dt =1$ and is taken to
be 
non vanishing during
a very small interval $\Delta t$.
Thus, the interaction term dominates the rest of
the Hamiltonian during $\Delta t$, and the time evolution
$
e^{-i P A}
$
leads to a correlated state: eigenstates of $A$ with eigenvalues $a_n$
are correlated to measuring device states in which the pointer is
shifted by these values $a_n$ (here and below we use units such that
$\hbar =1$). Thus in an  ideal measurements the final state
of the measuring device is  very simple related 
to the state of the measured system.
The properties of ideal measurements are:

a) The outcome of the measurement can only be one of the eigenvalues
$a_i$.

b) A particular outcome $a_i$ appears at random, with probability depending
only on the initial state of the measured system and
is independent of the details of the measurement.

c) The measurement leads to the (true or effective, depending on one's
preferred interpretation) collapse of the wave-function of the measured
system on the eigenstate
$\vert a_i\rangle $.
Subsequent ideal measurements of the same observable $A$ invariably yield 
the same eigenvalue $a_i$.

The opposite limit of extremely weak and long interaction 
is also clean enough to be called a measurement\cite{AV,AAV}. 
In such an adiabatic (or protective)
 measurement,
the coupling is very small:
$g(
t) = 1/T$ for most of the interaction time $T$ and
 $g(t)$ goes to zero gradually before and after the period
$T$. In order that the measurement
be as clean as possible, we also impose 
that:
the initial state of the measuring device is   such
that the momentum  $P$  is
bounded; that the momentum $P$ is a constant of motion of the  
whole  Hamiltonian
eq. (\ref{one}) (but we shall only 
 consider the simpler case where $H_{MD}$ vanishes);
and that the free Hamiltonian $H_0$  has non-degenerate eigenvalues $E_i$.
For $g(t)$ smooth enough we then obtain an adiabatic process in
which the system cannot make a transition from one energy eigenstate
to another, and, in the limit $T \rightarrow \infty$, the interaction
Hamiltonian          changes
 the energy eigenstate by an infinitesimal amount.
   If the initial state
of the system is an eigenstate $\vert E_i\rangle$ 
of $H_0$ then for any given value of
$P$, the energy of the eigenstate shifts by an infinitesimal amount
given by the first order perturbation theory:
$\delta E = \langle E_i \vert H_{int} \vert E_i  \rangle  =
\langle E_i \vert
A \vert E_i\rangle P/ T.
$ The corresponding time evolution $ e^{-i P \langle E_i \vert
A \vert E_i\rangle } $ shifts the
pointer by the expectation  value of 
$ A$ in the state $\vert E_i\rangle$. 
The main properties of adiabatic measurements are:

a) The outcome of the measurement can only be the expectation value 
$\langle A \rangle_i=\langle E_i\vert A\vert E_i\rangle$.
 
b) A particular outcome $\langle A \rangle_i$ 
appears at random, with a probability which
depends only on the initial state of the measured system and 
is independent of the details of the measurement.

c) The measurement leads to the collapse of the wave-function of the measured
system on the energy eigenstate $\vert E_i\rangle$ corresponding to the
observed expectation value $\langle A \rangle_i$\cite {Unruh}.
Subsequent adiabatic measurements of the same observable $A$ invariably yield 
the expectation value in the same eigenstate $\vert E_i\rangle$.

d) Simultaneous measurements of different observables yield the
expectation value in the same energy eigenstate $\vert E_i\rangle$.

The aim of the present letter is to consider 
measurements on
systems which evolve according to an
effective non hermitian Hamiltonian.
While ideal
(impulsive) measurements on such systems lead to no surprise (since in an
impulsive measurement the unperturbed Hamiltonian of the measured system
plays no role), adiabatic measurements yield as outcomes some new type
of values associated with the measured observable, namely the 
``weak values''\cite{AV2}.
Weak values 
where originally introduced in the context of the two state formalism
\cite{ABL,AV2,AV3,RA} wherein a system is described by two states, 
the usual one
$\vert \Psi_1 \rangle$ 
evolving towards the future from the initial time $t_1$, and 
a second state $\langle \Psi_2 \vert$ 
evolving towards the past from the final time $t_2$.
If at an intermediate time 
 a sufficiently weak measurement is carried out  
on such a pre- and post-selected system, 
the state of the measuring device after
the post-selection is shifted to
$
\Psi_{MD}(Q)\to \Psi_{MD}(Q-A_w) ,
$
where $A_w$ is the weak value of the observable $A$
\begin{equation}
A_w = { \langle{\Psi_2} \vert A \vert\Psi_1\rangle 
\over \langle{\Psi_2}\vert{\Psi_1}\rangle } .
\label{EE1}
\end{equation}
Note that weak values can take values which lie 
outside the range of eigenvalues
of $A$ and are in general complex. Their real and imaginary part
affect the position and momentum of the pointer respectively.
Weak values are associated with
two states which in the present context are the left and
right eigenstates of the effective Hamiltonian (see
below)\cite{AV-nyas}. 
The main
properties of adiabatic measurements carried out on a 
system evolving according to an
effective non hermitian Hamiltonian are:

a) The only possible outcomes of the measurement are the 
weak values  $A_w^i$
 corresponding to one of the pairs of states 
 $\langle \psi_i \vert \vert \phi_i \rangle$
associated with the non hermitian Hamiltonian.

b) A particular outcome $A_w^i$ appears at random, with a probability which
depends only on the initial state of the measured system and 
is independent of the details of the measurement.

c) The measurement leads to an effective collapse 
to the two-state vector $\langle \psi_i \vert \vert \phi_i \rangle$
corresponding to the
observed weak value $A_w^i$.
Subsequent adiabatic measurements of the same observable $A$ invariably yield 
the same weak value.

d) Simultaneous measurements of different observables yield the
weak values corresponding to the same two-state vector 
$\langle \psi_i \vert \vert \phi_i \rangle$.

Although the Hamiltonian of a quantum system is always a hermitian operator,
under suitable conditions a subsystem may evolve according to an effective
non hermitian Hamiltonian.  A well known case is the description of
metastable states \cite{WW}.
If the system is initially in the metastable state 
$\psi(0)$,
after a time $t$ it will be in the state $\psi(t) =
e^{-i H_{eff}t} \psi(0) + \ decay\ products$
where $H_{eff}$ is the effective non hermitian Hamiltonian. 
A celebrated example where this description 
has proved extremely useful is  the Kaon 
system.
Another case    in which  a system evolves 
according to an effective non hermitian
Hamiltonian is when it is coupled to a suitably pre- and post-selected
system\cite{AV-nyas}.
As an example, consider  a spin $1/2$ particle coupled 
to a pre- and post-selected system $S$ of large spin $N$
through the  Hamiltonian
\begin{equation}
H_0= \lambda {\bf S \cdot \sigma}.
\label{two}
\end{equation}
The large spin is pre-selected 
at $t_1$ to be in the state $|S_x {=} N\rangle$ and
post-selected to be at $t_2$ in the state 
$\langle S_y {=} N|$.
The coupling constant $\lambda$ is chosen in such a
way that the
interaction with our spin-1/2 particle  cannot
change significantly the two-state vector of the system $S$.
Indeed, the system with the spin $S$ can be considered as $N$ spin 1/2
particles all pre-selected in $|{\uparrow_x} \rangle$ state and
post-selected in $|{\uparrow_y} \rangle$ state. Since the strength of the
coupling to each spin 1/2 particle is $\lambda \ll 1$, 
during the time of the measurement their states cannot change
significantly. 
(However $\lambda N$ must be large so that the effective
Hamiltonian is significant.)
Thus, the forward evolving state
$|S_x {=} N\rangle$ and  the  backward evolving state $\langle S_y {=}
N|$ do not change significantly during the measuring process. 
Hence, effectively, the 
spin-1/2 particle is coupled to the weak value of 
{\bf $S$} 
\begin{equation}
{\bf S}_w = {{\langle S_y = N
|(S_x, S_y, S_z)  |S_x = N  \rangle} \over{\langle S_y = N
 |S_x = N  \rangle}} = (N, N, iN),  
\label{EE3}\end{equation}
and the effective non hermitian Hamiltonian is given by
\begin{equation}
H_{eff} = \lambda N (\sigma_x + \sigma_y +i \sigma_z).
\label{EE3B}
\end{equation}
The non hermiticity of $H_{eff}$ is due to the complexity of ${\bf S}_w$.
A detailed discussion of this example is given below.

Note that the effective non hermitian Hamiltonians
only arise due to a partial post-selection. In the spin example it 
only applies if the 
large spin is found in the state $\langle S_y = N
 |$. In the case of metastable states it only
 applies to the metastable states so long as they have not decayed.

We now analyze the general properties of a non hermitian Hamiltonian 
$H_{eff}$ which  
has non degenerate eigenvalues $\omega_i$.
In general the eigenvalues are complex.
Denote the eigenkets and the eigenbras of $H_{eff}$ by $\vert\phi_i\rangle$ 
and $\langle{\psi_i}\vert$:
\begin{equation}
H_{eff} \vert\phi_i\rangle = \omega_i \vert\phi_i\rangle, \quad \quad
\langle{\psi_i}\vert H_{eff} = \omega_i \langle{\psi_i}\vert.
\label{six}
\end{equation}
Contrary to the case where $H_{eff}$ is hermitian, the  $\vert\phi_i\rangle$ 
are not orthogonal to
each other, nor are the $\langle{\psi_i}\vert$, and furthermore $\vert
\psi_i\rangle\neq
\vert\phi_i\rangle$.
 However  the  $\vert\phi_i\rangle$
and $\langle{\psi_i}\vert$ 
each form a complete set, and they obey the mutual orthogonality condition
\begin{equation}
\langle{\psi_i}\vert{\phi_j}\rangle = \langle{\psi_i}
\vert{\phi_i}\rangle \delta_{i j},
\label{seven}
\end{equation}
which follows from subtracting the two identities
$
\langle{\psi_i} \vert H_{eff} \vert\phi_j\rangle = 
\omega_j  \langle{\psi_i}\vert{\phi_j}\rangle$,
$\langle{\psi_i}\vert H_{eff} \vert\phi_j\rangle = 
\omega_i  \langle{\psi_i}\vert{\phi_j}\rangle$
for $i \neq j$. Eq. (\ref{seven}) enables us to rewrite $H_{eff}$ as
\begin{equation}
H_{eff} = \sum_i \omega_i  { \vert\phi_i\rangle \langle{\psi_i}\vert
 \over \langle{\psi_i}\vert{\phi_i}\rangle },
\label{nine}
\end{equation}
which generalizes the diagonalization of hermitian operators.
The  eigenkets of $H_{eff}$ are the natural basis in which to decompose
a forward evolving state $\vert\Phi\rangle$. Indeed, 
using the decomposition of unity
$ I = \sum_i {\vert\phi_i\rangle\langle{\psi_i}\vert \over
\langle{\psi_i}\vert{\phi_i}\rangle }$
one obtains
\begin{equation}
\vert\Phi\rangle
= \sum_i {\langle{\psi_i}\vert{\Phi} \rangle\over
\langle{\psi_i}\vert{\phi_i}\rangle }  \vert\phi_i\rangle
=\sum_i \alpha_i \vert\phi_i\rangle 
\label{EE6}\end{equation}
(On the other hand a backward evolving state should be decomposed
into the eigenbras of $H_{eff}$ as
$\langle{\Psi}\vert = \sum_i \beta_i \langle{\psi_i}\vert$).
The formal solution of the Schr{\" o}dinger's equation with the
effective
Hamiltonian $H_{eff}$ is:
\begin{equation}
\vert\Phi(t)\rangle =
e^{-i H_{eff}t}\vert\Phi\rangle = 
\sum_i \alpha_i e^{-i\omega_i t} \vert\phi_i\rangle
\label{EE7}\end{equation}
Note that the norm $\cal N$ of $\vert\Phi(t)\rangle$ is not equal
to 1 but is time dependent. 
Formally, there are two causes for not conserving the norm in time
evolution due to the effective Hamiltonian. The first is that
the eigenvalues $\omega_i$ may be complex. The second is that
the eigenkets are not necessarily orthogonal. 
This non conservation of probability by non hermitian Hamiltonians
has a natural interpretation when one recalls that  we are
describing partially 
post-selected systems.
In the case of metastable 
states 
${\cal N}(t)$ is the probability for the states not to have decayed.
In the spin example
${\cal N}(t)$  describes corrections to the 
probability of finding the state $\langle S_y = N|$.

Let us illustrate this general formalism by considering the Kaon system.
The two eigenkets of the effective Hamiltonian
are traditionally denoted $\vert K_L\rangle$ and $\vert K_S\rangle$.
Similarly, one can define the eigenbras of the effective Kaon
Hamiltonian 
$
\langle{K_L'}\vert $ and 
$
\langle{K_S'}\vert$. 
The particular features of non hermitian Hamiltonians are controlled
by the CP violation parameter $\epsilon\simeq 10^{-3}$. 
The non orthogonality of the eigenkets is
$\langle {K_S}\vert{K_L}\rangle = O(\epsilon) $ 
and the non equality of the
right
and left eigenstates is $\langle{K_L'}\vert {K_L}\rangle = 1-O(\epsilon^2)$.
In view of the smallness of $\epsilon$  
the adiabatic measurements which we propose below may be difficult to 
implement in the Kaon system. However, other metastable systems
may display much stronger non orthogonality and be more amenable to 
experiment.

In the spin example, the effective Hamiltonian eq. (\ref{EE3B}) has two
eigenvalues $+\lambda N$ and
$-\lambda N$ with eigenkets (bras) $\vert\uparrow_x\rangle$
($\langle{\uparrow_y}\vert$) and $\vert\downarrow_y\rangle$ 
($\langle{\downarrow_x}\vert$) respectively.
Thus, $H_{eff}$ can be rewritten as
\begin{equation}
H_{eff} = \lambda N {\vert\uparrow_x\rangle\langle{\uparrow_y}\vert 
\over \langle{\uparrow_y}\vert{\uparrow_x}\rangle}
-
\lambda N {\vert\downarrow_y\rangle\langle{\downarrow_x}\vert 
\over \langle{\downarrow_x}\vert{\downarrow_y}\rangle}.
\end{equation}
In this example the eigenkets and eigenbras associated with the same
eigenvalue
are very different. Thus, weak values associated with these two states 
can have surprising values. For example,
$\langle{\downarrow_x} \vert\sigma_z \vert\downarrow_y\rangle /
\langle{\downarrow_x}\vert{\downarrow_y}\rangle = -i$, which is pure imaginary
and 
$\langle \downarrow_x\vert( \sigma_x +\sigma_y)/\sqrt{2} \vert\downarrow_y
\rangle /
\langle{\downarrow_x}\vert{\downarrow_y}\rangle= -\sqrt{2}$,
which lies outside the range of eigenvalues of ${\bf \sigma \cdot n}$.

We are now ready to discuss  adiabatic  measurements performed on a 
system evolving according to $H_{eff}$.
The Hamiltonian describing such a measurement is
given by eq. (\ref{one}) with $H_0$ replaced by $H_{eff}$.
The coupling parameter
$g(t)$ equals $1/T$ for most of the interaction 
time $T$ and goes to zero gradually before and after the period
$T$.  In order that the measurement be as clean as possible we 
also impose that:
$H_{eff}$  has non degenerate eigenvalues; that
the initial state of the measuring device is  such
that the momentum  $P$  is
bounded; and that the momentum $P$ is a constant of motion of the  
whole  Hamiltonian
eq. (\ref{one}).
For $g(t)$ smooth enough, and in the limit
$T \rightarrow \infty$,  we obtain once more an adiabatic process 
such that if the system is initially in an eigenket $\vert\phi_i\rangle$,
it will still be in the same eigenket after the measurement. Furthermore,
in this limit,
the interaction
Hamiltonian changes the eigenket during the interaction
by an infinitesimal amount.

If we take the initial state
of the system to be an eigenket  $\vert\phi_i\rangle$, then for any given value
 of
$P$, the eigenvalue of the eigenstate shifts by an infinitesimal amount
which can be obtained using first order perturbation theory as follows. The 
perturbed eigenstates are solutions of
\begin{eqnarray}
&\left( 
H_{eff} + {P \over T} A \right)
\left( \vert\phi_i\rangle + \sum_{j\neq i} c_{ij} \vert\phi_j\rangle \right)
=& 
\nonumber\\
&
\left( \omega_i + \delta\omega_i\right) 
\left( \vert\phi_i\rangle + \sum_{j\neq i} c_{ij} \vert\phi_j\rangle
\right).
&
\label{A1}
\end{eqnarray}
Taking the scalar product with $\langle{\psi_i}\vert$, 
to first order in $P/T$ one
obtains
\begin{equation}
\delta\omega_i =
{P \over T} { \langle{\psi_i} \vert A \vert\phi_i\rangle \over 
\langle{\psi_i}\vert{\phi_i}\rangle } = {P \over T}A^i_w .
\label{A2}
\end{equation}
Thus the state of the measuring device after the measurement
is  shifted,
$
\Psi_{MD}(Q) \to \Psi_{MD}(Q - A^i_w)
$, and if the initial wave function of $\Psi_{MD}$ is sufficiently
peaked in $Q$, the reading of the measuring device yields the weak
value of $A$.

It is instructive to consider the case when the 
initial state is not an eigenket of $H_{eff}$.
 The initial state should then be decomposed into
a superposition of eigenkets $|\Phi\rangle = \Sigma_i
 \alpha_i |\phi_i\rangle
$
and its time evolution, up to normalization, will be given by
\begin{equation}
\vert \Phi\rangle \psi_{MD}(Q)
\to
 \Sigma_i \alpha_i e^{-i\omega_i T}|
 \phi_i\rangle \psi_{MD} (Q- A_w^i ).
\label{E3}\end{equation}
The state of the measuring device is amplified to a macroscopically
distinguishable situation. Then, effectively, a 
collapse takes place to  the reading of one of the  weak values
$A^i_w$ with the relative probabilities given by $|\alpha_i\vert^2
e^{ 2 Im(\omega_i) T}$.
We call the collapse effective because it only occurs under the
condition that a partial post-selection is  realized.
 A
subsequent adiabatic  measurement
of another observable $B$
will yield
the weak value corresponding to the same two-state 
$\langle{\psi_i}\vert    \vert\phi_i\rangle$.
Alternatively, one can carry out the measurements of $A$ and $B$
simultaneously. This can always be done by increasing the duration
$T$ of 
the measurement so that the 
interaction $(P_1 A + P_2 B)/T$ remains  a small perturbation. 
Thus, given a sufficiently long time $T$, one can obtain
reliable
measurements of any set of observables by making  measuring devices
interact adiabatically with a single quantum system. 
However it should be noted that in any realistic implementation we
will need ensembles of systems and measuring devices since both in the
case of metastable states and in the spin example the probability of a
successful partial post-selection (which gives rise to the effective
non hermitian Hamiltonian) is very small. 
Indeed, the adiabatic measurement
will only be successful if the metastable states do not decay during
the measurement, or if the     spin $S$ is found in the state $\vert S_y
  =N\rangle$.  Nevertheless, there is a non-zero probability that the first
run with a single system and a single set of measuring devices will
yield the desired outcomes.

Our general discussion was carried out for a system evolving according
to an arbitrary effective non hermitian Hamiltonian. The spin example
presented above is amenable to exact treatment and one can
investigate in this case in what limit the effective non hermitian
Hamiltonian describes adequately the evolution of the spin $1/2$
particle.
We recall that the effective 
Hamiltonian eq(\ref{EE3B}) has two eigenkets $\vert{\uparrow_x}\rangle$
and $\vert{\downarrow_y}\rangle$.  
That $\vert{\uparrow_x}\rangle$ should be an eigenket 
is  easily be seen by noting that the initial state
$\vert S_x =N\rangle\vert{\uparrow_x}\rangle$ 
is an eigenstate of the free Hamiltonian 
$H_0 = \lambda {\bf S  \cdot \sigma}$. 
That $\vert{\downarrow_y}\rangle$ is an eigenket is a 
nontrivial prediction which 
can be checked by calculating the probability for the small
spin, initially in the state $\vert{\downarrow_y}\rangle$, to  be in 
the state $\vert{\uparrow_y}\rangle$ at an intermediate time.
One finds that this probability is proportional to $1/N^2$,
thereby confirming that it is indeed an eigenket in the limit
of large $N$.

If the initial state of the small spin is $\vert{\downarrow_y}\rangle$, and
an adiabatic measurement of $\sigma_\xi = {\bf \sigma \cdot \hat{\xi}}$ is
carried out the 
eigenket $\vert{\downarrow_y}\rangle$ should be unaffected by the measurement,
and the pointer of the measuring device should be shifted by
$(\sigma_\xi)_w =
{\langle{\downarrow_x}\vert \sigma_\xi \vert{\downarrow_y}\rangle \over
\langle{\downarrow_x}\vert{\downarrow_y}\rangle}$.
In order to verify this we considered the particular case when
$\hat   \xi = \hat   x$ whereupon  the analysis simplifies
considerably  since only the states 
with $J_x=S_x+\sigma_x=N+1/2,\  N-1/2, \  N-3/2$  come up in the calculation.
Thus, we took the Hamiltonian to be
$H= \lambda  {\bf S \cdot \sigma} +{P\over T} \sigma_x$ 
during the interval $t_1<t<t_2=t_1+T$, with the
initial state $\vert S_x = N\rangle\vert{\downarrow_y}\rangle$
and the final state of the large spin post-selected 
to be $\langle{S_y =N}\vert$.
Taking the measuring device to be in the momentum eigenstate $P$,
one finds that after the post-selection, 
at $t=t_2$, the state of the small spin plus measuring device is
$\vert{\downarrow_y}\rangle e^{i P /2} +\ error\ terms$.
The error terms are either of the form
$ \vert{\downarrow_y}\rangle e^{-i P /2}$ corresponding to a pointer shifted in
the 
wrong direction, or  of the form
$f(P)\vert{\uparrow_y}\rangle$ corresponding to the spin not having remained
in the state $ \vert{\downarrow_y}\rangle$. The norm of the error terms is 
proportional to $1/N$ and
in the limit of large $N$ they  can be neglected. One then finds
that
after and during the measurement the spin is still in
the eigenket  $\vert{\downarrow_y}\rangle$
and that the pointer of the measuring device is shifted by the weak value
$(\sigma_x)_w = { \langle{\downarrow_x} \vert
\sigma_x \vert{\downarrow_y}\rangle \over
\langle {\downarrow_x}\vert  {\downarrow_y}\rangle } = -1$.
Thus we confirm that in the limit of large $N$, the evolution is
given by the effective non hermitian Hamiltonian.

In this letter we have analyzed adiabatic measurements on systems
which evolve according to an effective non hermitian Hamiltonian.
The effective Hamiltonian only arises when a partial
post-selection is realized. For an adiabatic measurement to yield 
a significantly unusual result, the non hermiticity of the Hamiltonian
must be large, and in such cases the probability of a successful
partial
post-selection is very small.
There is however a reasonable hope of performing such a measurement in
a real laboratory.
It is conceivable to build an
experiment in which the measuring device is a particular degree of
freedom of the measured particle itself, and in this case the
post-selection process is particularly simple\cite{RSH}.

This research was supported in part by  grant 614/95 of the 
Basic
Research Foundation (administered by the Israel Academy of Sciences and
Humanities), by ONR grant no.R\&T 3124141 and
by NSF grant PHY-9321992. One of us (J. T.) would like to 
acknowledge the support of the Fetzer Institute.

\end{multicols}

\end{document}